# Brain-Computer Interfaces and the Dangers of Neurocapitalism

Srdjan Lesaja[1] and Xavier-Lewis Palmer[2]

[1] Virginia Commonwealth University, Richmond VA, 23284, USA
[2] Old Dominion University. Norfolk, VA 23529, USA
`slesaja@vcu.edu`

**Abstract.** We review how existing trends are relevant to the discussion of brain-computer interfaces and the data they would generate. Then, we posit how the commerce of neural data, dubbed Neurocapitalism, could be impacted by the maturation of brain-computer interface technology. We explore how this could pose fundamental changes to our way of interacting, as well as our sense of autonomy and identity. Because of the power inherent in the technology, and its potentially ruinous abuses, action must be taken before the appearance of the technology, and not come as a reaction to it. The widespread adoption of brain-computer interface technology will certainly change our way of life. Whether it is changed for the better or worse, depends on how well we prepare for its arrival.

## 1 Introduction

### 1.1 Relevant Concepts

Here, we briefly introduce trends and phenomena that are relevant to our forthcoming argument, and which together form the context and climate in which the argument is levied.

**Digital Space.** The last two decades have witnessed a first for humankind: the emergence of a truly new space. Once only a supplement to the physical world, the network density of the Internet has surpassed some threshold and gestated into a wholly separate digital space. Here, entire ecosystems and communities are formed, without any tether to a physical counterpart.

We refer to the Internet, and sum of all network-connected computers, devices, and storage, as the Digital Space. We consider it a separate space because it does not follow the same laws as physical space and requires the cultivation of a separate conceptual reference frame. Familiar concepts like distance, accountability, privacy, and personal identity, have entirely different meanings in the Digital Space. With the evolution of the Digital Space, we are feeling the growing pains of developing new conceptual frames. We argue that the mismatch of applying physical space concepts to the digital space is one of the issues at the heart of the exploitation discussed. We cannot apply our existing mental models to this new space.

**Surveillance Capitalism.** In our current economic climate, the advent of a new technological paradigm is followed closely by a market seeking to monetize a new theater of resources. Social networks, smartphones, Internet of Things (IoT) devices, increased computational speed, decreased cost of digital storage; these have all played a role in connecting us on an unprecedented scale. They have also enabled the rise of Surveillance Capitalism. Every action in the digital space is mediated by a computer-executed command, and nearly all can thus be logged, potentially forever. User data like personal information, browsing habits, purchases, and behavioral patterns, are all mined, aggregated, and brokered. Shoshana Zuboff and others have extensively detailed the origins and detrimental effects of this user exploitation [1]. While many reasons contribute to this problem, we believe a key contributor is the users' lack of full comprehension of the consequences of their online actions.

**Nudging and Consent.** Even where limited options exist for control of what user data is stored, those options are often hard to find, and the default setting favors the gathering of user data. Some companies have abused Terms and Conditions that users must agree to in means that have encouraged irresponsible acceptances [2-4]. They use vague language and are so excessive in length that a layperson could not be reasonably expected to have read or understood them[2-4]. Users are rarely if ever educated on what data will be collected, and how that data will be used [2-4]. The burden of self-elucidation is left to the user, and this burden is too great [2-4].

Users, especially ones that did not grow up in the information age, are still largely ignorant of the capabilities allowed by the digital space, and so fall back to their understanding of physical space. But, filling out a physical form with your name and address that will sit in a file cabinet is very different from providing that same information online. There is ignorance about the ease and extent of data acquisition, and the persistence of acquired data. Additionally, there is ignorance about the value of this personal data.

Companies prey on user ignorance. Users are unethically nudged into agreeing to terms they do not fully understand under the guise of consent. We argue that this cannot be considered consent because it is not informed consent and is instead a form of coercion.

**Lie of Opting Out.** An argument given in support of Surveillance Capitalism is that if users do not want to agree to the terms, they have a choice. They can simply choose to not use the service. While technically true, this has become a practical impossibility. Many large corporations that engage in the surveillance economy have worked to make their platforms so pervasive that to avoid using them takes sometimes overwhelming effort and would put people who make this choice at a significant disadvantage.

There is a case to be made that some of the services are so integral that they should be regulated as utilities. Unfortunately, since there is still debate as to whether internet access itself should be a utility, it seems unlikely that derivative and adjacent services such as email or GPS will be seriously considered soon. Whether or not such tracking items will be does not change the fact that many in society could not function normally today without utilizing email, GPS, or similar technologies. Their use relies on the agreement of one-sided terms, dictated by a company. These terms often offer very little regulatory oversight and even less in the way of required user protections, at least in the United States.

**Machine Learning Advancements.** Surveillance Capitalism has developed hand in hand with information age and Big Data trends. Increased access to data and decreased data storage costs meant choices no longer had to be made about which data to store. All information could find a use, or a buyer. Increased computational speed and distributed computing enabled mathematical data analysis techniques that were previously too computationally intensive. The coupling of large amounts of available data and newfound ability to handle it saw a renaissance in machine learning and artificial intelligence applications.

Machine learning algorithms make possible many familiar computer interactions: including speech to text, context-aware searching, media suggestion, GPS guidance, email auto-response and phrase completion, and psychometrics. For many businesses that profit from user surveillance, of the data models previously mentioned, all have been built on user data and would not be nearly as effective without them. Users, however, are not fairly compensated for the vital role their data played in the advancement of this field.

**Personal (Behavior) Models.** Machine learning techniques began with population-level models, but as the field has progressed, models are becoming person-level almost at default. Data models have been personalized to the user's voice, their face or fingerprint, their browsing or purchase habits, the people that they associate with, or the places they travel. We refer to a data model, or simply a model, as any machine learning algorithm designed to take "natural", non-structured data (such as speech, browsing data, social network connections, or content of text), convert this data to computable input features, and use these features to classify, infer, or predict subsequent outcomes. The accuracy of a model improves with the amount of data utilized for training, and their inputs can be the outputs of other models, creating self-reinforcing constellations of machine learning.

The final algorithm listed in the previous section, psychometrics, deserves special consideration. Psychometrics is a field concerned with reducing mental states, behaviors, beliefs, or traits. In this context, we speak of these being estimated from user data. Whether collected firsthand or bought, this data is used to create a personalized model of behaviors, traits, temperaments, beliefs for all of these. Often the creation of these personalized models is advertised as a boon to the user, creating better search results, or purchase recommendations. Not mentioned nearly as often is how these same methods can be abused.

**Data Violence.** Abuse of personal models is always unethical, but as they are being applied to more and more areas, we are seeing detrimental effects beyond exploitation or coercion. The misapplication of these algorithms can cause serious harm. Models are only as insightful as their architecture allows, and in every case, can only ever imply correlation, which is obscenely often confused for causality. Two variables can be highly correlated, but have no causal bearing on one another, as spurious correlations are all around us [5].

If a model carelessly employs demographic information as an input, that model runs a high risk of being discriminatory, potentially creating both a foundation and context for violence against a group [6-8]. When model predictions weigh in on the livelihoods and lives of people, any such risk is unacceptable [6-8]. Currently, not enough care is being taken and there is not enough oversight to ensure discriminatory or similar input is not occurring [6-8]. As many have written, the digital space is a new theater for discrimination, and while they should have ushered greater equality, models are too often weapons [6-8]. This concept is further developed by Dr Anna Hoffman, professor and researcher of data's ethical and techno-cultural intersections ; we recommend that readers refer to works of hers such as "Data, Technology, and Gender. Spaces for the Future", Data Violence and How Bad Engineering Choices Can Damage Society", and Where fairness fails: data, algorithms, and the limits of antidiscrimination discourse" [6-8].

**Tailored Communication.** On the other side of the model abuse spectrum, there is a more subtle but no less insidious application. We mentioned behavior models being used for personalization of advertisements. Before, a product may have been marketed two different ways to two separate demographics. Now, with person-level personality models and social media advertising, a product can be marketed differently to everyone.

At first glance that may seem innocuous, but in some sense, advertising has become the study of how to best craft a message so that it is most likely to be received. The idea of personally tailored messaging is very powerful. The idea of being able to purchase the ability to deliver a tailored message to an entire population should be cause for alarm.

In 2013, Kosinski et. al. showed "that easily accessible digital records of behavior, Facebook Likes, can be used to automatically and accurately predict a range of highly sensitive personal attributes including: sexual orientation, ethnicity, religious and political views, personality traits, intelligence, happiness, use of addictive substances, parental separation, age, and gender" [9]. Therein, Kosinski et al. also warn about the potential for abuse of behavioral targeting. Several years later, these same techniques were used by Cambridge Analytica when they were employed by the Donald Trump presidential campaign in 2016 to deliver tailored advertisements to Facebook users[9].

Consider any message to be the contents of a package, and the way the message is relayed the packages wrapping. If someone learns what wrapping someone is most likely to accept, then they increase the chance of delivery, regardless of what the contents may be. Tailored messaging and mass persuasion are already commonplace, increasing in efficacy with increasing availability of data. As these tools become more effective, they place individuals at progressively greater risk.

**Appealing to the Subconscious.** Unfortunately, tailored messaging is often effective even if one is aware of its existence. This is because it appeals to ingrained behaviors and habits that are subconscious. By its nature, a bias is a blind spot that someone is unaware of, but it can be exploited. Appealing to the subconscious has been used from the very beginning of modern advertising. Until recently, however, the mechanism has followed a wide and stochastic loop. The success of such an appeal was measured in downstream outcomes (sales), or perhaps more directly in a controlled setting (focus group). With the increase in wearables and IoT devices in tandem with the surveillance economy and personal behavior models, this loop threatens to become much tighter, and much less stochastic. Physiological responses can be quite accurate indicators of conscious and subconscious reactions to presented stimuli. One can imagine a scenario in which a heart rate monitor someone is wearing detects an increase in response to an ad that was seen. This adds a very valuable feedback input that a model can use to achieve better accuracy and a more complete picture.

As we continue to become more intertwined with technology, the capability of these models to receive direct cause and effect information will increase, and perhaps the loop will eventually even approach real-time.

**Brain-Computer Interfaces.** Ultimately, the goal of psychometric analysis is to assess what someone is thinking. The field of Brain-Computer Interfaces (BCI) concerns itself with meaningfully measuring brain states with the goal of interfacing with a computer. Interfaces bridge us and the digital realm, allowing for the extraction and imprint from one side to the other. Briefly defined, they are any device designed to sense the physiological brain state. The goal of any such sensing modalities is to determine the patterns of activity of neurons. How closely these measurements of neuronal activity can be decoded to reflect actual thought is an open question, but there is value in addressing problems with BCIs before this is certain, for both benefits of the technology and the deterrence of misuses. BCI technology is still largely confined to academic laboratories, and impractical for robust widespread use. But that will not always be the case. When it does leave the lab, it will represent yet another step in the further externalization of our internal states. As the field develops, it will no doubt interact with all the phenomena outlined above, shape and be shaped by them. We must act now to prevent this escalation following the same trends that doom the future of responsible use.

## 2  Neurocapitalism

**Definition.** Here we use the term Neurocapitalism to refer to an economic subsystem surrounding the collection, commodification, use, and brokerage of any measure that serves as proxy for a neural state. Put another way, the resource of Neurocapitalism is Information About People's Thoughts [10].

Neurocapitalism has a multitude of commonalities with the following terms of which it is a subset: information economy and Surveillance Capitalism. In talking about such, it is important to make a distinction between Information About People and Information About People's Thoughts. The resource of Neurocapitalism is neither general information, demographic information, nor information about a person like where they reside or their income. It is information about thoughts, in all their forms, such as intents, beliefs, traits, actions, temperaments or choices. Furthermore, we say information about thoughts, rather than thoughts themselves, because they cannot be measured directly. However, downstream effects of thoughts, such as heart rate or browsing data, can be measured. These measures are what we refer to as neural data. Neural data can then be used to infer the root thoughts that generated them.

**Producer-Consumer Transaction Loop and Fair Compensation.** Neural data is a fundamental component in the production of many machine learning products. Often, the quality of these products is recursively tied to user data. The more users of an application, the more data can be collected, to better inform the models, which provide better prediction and function, which in turn attracts more users.

We argue that the nature of this relationship is why the current state of Neurocapitalism requires serious attention, as it is based on flawed assumptions. The current paradigm frames the user of a technology primarily as the consumer. The company owns capital, often in the form of technology that uses data models. The user is allowed use of this technology, sometimes for free, and in exchange for the use and sometimes rights to sell their usage data of the technology.

At first glance this may seem like a fair transaction, we believe it is skewed toward the advantage of the company. The user only benefits from the technology while they are using it. The company retains their neural data, and the value added by it to their models, indefinitely. Because of the persistence of data in the digital space, the company can potentially profit off a user's data long after they have discontinued the use of their technology. In fact, for some general data models, their nature is such that once user data has been added to improve the model, it cannot be removed. Thus, the user has permanently improved the company product, for a temporary benefit. In an interaction whose terms are entirely dictated by the company.

But it is far from a one-way transaction. The company is also a consumer of the user's neural data. The user is producing neural data. If we take the view that the company is producing a commodity, and "mining" this neural data, then the user should be considered a laborer, and compensated fairly for their part in the production.

If we take the view that the user is the entity producing the commodity, then they should be fairly compensated for its use. Usage of the technology is not a fair compensation as the user of technology essentially becomes an "infinite consultant" without the benefit of consistent and proportional payment for their education, time, and insight. Who would have better insight on the customer that is you, than you? In this way the end user is strangely also the initial producer. What company would like to take on infinite consultants and could fairly compensate them?

**Uniqueness of Neural Data.** Of the two views mentioned above, the authors believe the second is more accurate. There is an interesting property of neural data as a resource. Namely novelty. The amount of neural data that a user produces is not a linear function with respect to time. As time increases, the amount of new neural data that is produced has diminishing returns, and the amount of redundant data increases. Tautologically, every use can only ever produce one unit of *Information About A Person's Thoughts*; specifically, their own. An analogy might be a painting. Each user has one canvas. The more time they spend on the painting, the more detail it will have, but they can only ever produce one painting. From the perspective of the resource consumer (data aggregators, machine learning models), quantity has a quality all its own. E.g. Several moderately detailed paintings would be more valuable than one hyper-detailed painting. The property supports the view of each user being a producer, rather than a laborer.

This producer is one of a unique resource, which they can only ever hope to produce one of, regardless of the amount of time input. They are generating priceless artifacts.

## 2.1 Neurocapitalism and Brain-Computer Interfaces

Here we expand more on the state of BCI, what is possible now, what will likely be possible in the near future, and how this would interact with Neurocapitalism trends.

**BCI Primer.** BCIs attempt to sense the physiological brain activity. There are several types of BCI, and here are some attributes commonly used to differentiate them. Sensing modality: the underlying physical phenomenon that is being measured. Portability: how small a BCI can be and still sense accurately. This is especially important for non-clinical BCI. Time resolution: how little time needs to pass between each measurement. This is important for BCIs that require tight feedback loops, like BCIs that aid in balance rehabilitation. Spatial resolution: how accurately can the source of the measurement be pinpointed. Invasiveness: how time-consuming, difficult, or dangerous is it to place or remove the BCI. As a generality, more invasive BCI have better time resolution, spatial resolution, or both, but at the risk of requiring surgery.

A neuron firing produces an electric effect, and the most common forms of BCI are ones designed to sense measure electrical activity. Other modalities exist, however. Functional magnetic resonance imaging (fMRI) measures blood flow to brain regions. It has good spatial resolution and moderate time resolution but requires an MRI machine so is prohibitively non-portable for most any aim that is not clinical or research focused. Functional near-infrared spectroscopy (fNIRS) often measures changes in hemoglobin by shining an LED and measuring how much light of a specific wavelength was absorbed. It is a method similar to how wrist-worn heart rate trackers work. The modality is portable, non-invasive, has good time resolution, and moderate spatial resolution. fNIRS and EEG comprise the two modalities found in consumer-grade BCIs. Magnetoencephalography measures changes to the magnetic field made by the electrical activity of the brain. It requires a magnetically shielded room and an apparatus at least as large as an MRI machine. It is non-invasive, but prohibitively non-portable.

BCIs that measure electric activity can be grouped into their own category. The most common and oldest of the sensor types, were first used in the 1920s. These all share the same modality, measuring the electric field disturbance caused by neurons firing. What changes is size of the sensor (called an electrode), and its proximity to the brain, and we introduce a non-exhaustive list here.

A micro-electrode array is a tightly packed grid approximately, 1cm by 1cm, of tiny electrodes that is placed on the brain and whose electrodes penetrate slightly into the cortex. They have extraordinary spatial and time resolution, able to pick up single neuron firing. However, they are very limited on the area they span, and also extremely invasive.

Electrocorticography (ECoG) is an array of larger, disc-like electrodes placed on the surface of the brain. They sense aggregate activity of a larger set of neurons than a micro-electrode array, at the trade-off of being able to cover a much larger brain area. They also require surgery.

Electroencephalography (EEG) are electrodes placed on a hairless scalp. The time resolution of EEG is very good, but the spatial resolution is significantly affected as the electrical field disturbances diffuse significantly traveling through the skull. It is most directly comparable to FNIRS.

**Challenges of BCI.** Neuronal firing patterns are thoughts. Thoughts lead to actions. Repeated actions are behaviors. Behaviors are often manifestations of traits or beliefs. If we assume that neuronal firing patterns are thoughts, and that these thoughts lead to actions, we come across a problem with the use of BCIs. Afterall, repeated actions are behaviors and behaviors can manifest from traits or beliefs. Currently, by necessity, most neural data is measured with some action taken. If a stimulus produces a thought, but that thought does not manifest in some externalized action, then it cannot be measured. BCI has the potential to overcome that limitation. If neuronal firing patterns are measured directly, and those patterns are correctly mapped to thoughts, then it would not require a stimulus to produce an action in order for its reaction to be measured.

This motivates the two big challenges facing the field of BCI: feature extraction and feature translation. First, feature extraction is the question of accurately acquiring meaningful information. In a stadium full of people, there are some conversations happening, and then there is also a lot of noise. The challenge is how to differentiate between the two, especially if you do not know the language being spoken in the stadium. The last part here, alludes to the second big challenge: feature translation. Being able to map sensor measurements accurately to thoughts or intentions is very much akin to having to learn a language. The challenge is further complicated by the fact that almost every brain will have its own unique language.

**Current and Future Prospects.** The aforementioned two big challenges are two very hard problems that do not always lend themselves to being solved incrementally by being separated into smaller problems. Despite these hurdles to overcome, we believe that BCI will follow a similar trajectory of adoption as smartphones and wearable computing; and that it is not a question of if, but when.

This year, invasive ECoG BCI was shown to be able to decode a limited set of sentences solely from brain activity with 93% accuracy, and another was able to restore both limited movement and a sense of touch to a paralyzed limb following a spinal cord injury [11-12]. Elon Musk's company, Neuralink, has announced their own invasive BCI [13]. For non-invasive BCI, Facebook has announced plans to develop the technology, and has acquired several BCI and wearable startups in recent years [14]. An increasing number of classrooms in China have students wearing EEG trackers [15] Perhaps this is the most telling evidence that those interested in surveillance have noticed the potential of BCI.

## 3   Dangerous Consequences

The authors of this paper believe in technology's ability to do good and improve the lives of people. We are certainly not suggesting that no good has come of all the technologies discussed here, or that the world would be better off without them. Our view is that all technology is an instrument, and its ability to create or destroy, empower, or oppress is dictated by those that wield the technology, not the technology itself.

Unfortunately, it is very often much easier to use technology to exploit. Cooperation requires consensus, while exploitation requires relatively few, sometimes just one. It is with this in mind that several potential dangers that could arise if current trends continue and this fledgling technology matures are presented. Not because we believe there is no potential benefit, but precisely because we hope to see the dangers sidestepped, so that its full potential can be realized.

**Thought Monitoring and Policing.** Our minds have always been a bastion. A safe haven that could not be intruded upon. If BCI is capable of measuring a mental reaction to a stimulus, whether it be conscious or unconscious, then if strict rights to privacy aren't ensured, it could easily lead to thought monitoring. One could imagine a scenario in which a negative reaction to a superior is met with reprimand in a military setting. Or level of focus on a task in a work or school setting becomes a performance metric. Thoughts of anger or violence punished preemptive to actions. Could it become a world where the lines of consequence between thought and action are blurred? Even in a scenario where wearing of a BCI is voluntary and not legal mandate or contractual obligation, access to this neural data could be used to extort, coerce, or blackmail. What would it do fundamentally to our neurological development and psyche if we were no longer free to think our thoughts without fear of potential consequence, or fear of being judged on and responsible for our subconscious reactions?

**Othering.** Looking on the other side of the coin from the personal experience of thought monitoring, from the societal perspective, BCI technology existing without protections against intolerance and prejudice would be a weapon used to discriminate against disenfranchised groups. Without a culture of acceptance, and firm social and governance support structures, the technology could be potentially disastrous for human rights. Imagine such a technology existing during World War II and the Holocaust, or during the Red Scare, or the Civil Rights movement. The (potentially false) promise of measuring someone's allegiances is a seductive proposition. Any hegemony would require significant discipline and oversight not to succumb to the allure. Are we prepared as a society to not abuse this power when the ability to do so is trivial? Will that still be true in times when we feel our way of life is threatened?

**Enabling Addictions.** The same mechanism of measuring mental reactions to stimuli could be used to exacerbate some of the most detrimental aspects of advertising. Even without BCIs, it is the current modus operandi to appeal to people's biological and social drivers to help sell a product such as a slow motion food advertisement designed to make our mouths water.

Personalized advertising is already here, but what if the commerce of Neurocapitalism and BCI technology gives companies access to user reactions to advertisements? The technology would reveal the intensity of effect that a liquor ad has on a recovering alcoholic, for example. Changing unhealthy behaviors is extremely challenging. The biofeedback that BCI could provide could prove a useful tool to help people in changing these behaviors. However, knowledge of a temptation in the hands of advertisers could be used alarmingly effectively to sell a product. Current data suggests that we cannot trust companies to behave responsibly, and refrain from acting against peoples' best interest.

**Tailored messaging, Trust, and Identity.** Beyond advertising, we have already touched on the danger that personalized behavior modeling and neural data, especially with BCI data incorporated, could pose a fundamental change in the way we interact with messages and information in general.

We each have preferences, boundaries, triggers, and tendencies. We are able to convey these, and learn them of one another, usually through continued interaction. This knowledge is powerful as it allows someone both to communicate more effectively with us, and inherently also gives the power to manipulate us. Currently, we control these keys to communication because we choose with whom we interact, what information we share, and to an extent, what behaviors we exhibit. Though this statement is decreasingly true for interactions in the digital space. A fully realized person-model represents giving over control of these keys to a separate entity. If they are a commodity to be sold, then we would no longer have agency of choice as to who possesses or has access to them. How could we be sure that information we are receiving was not designed to manipulate us? What would this do to our ability to judge, or our sense of trust?

In other words, we stand the chance to lose the ability to curate our expressions while also losing the ability to freely perceive. Instead someone curates our experience for us. This would be ruinous for our conceptions of healthy relationships in society, and building our sense of meaning, especially if we desire to function with a sense of autonomy.

**Empowering Hegemonies.** In short, BCIs are likely to help cement existing hegemonies. Considering the brokering of neural data, personal behavior models, and targeted advertising, the addition of BCI technology will only make mass persuasion campaigns like that of Cambridge Analytica more frequent and effective.

In theory, it would be possible to craft a separate personalized message for everyone for whom there exists a personality model, but it would be expensive. A mechanism for those with extreme wealth or power to consolidate greater amounts. Neurocapitalism would create a more direct link between wealth and public influence than already exists. The nature of that influence might be more dangerous than we estimate. The most dangerous kind of manipulation is imperceptible. A gentle steering in a direction of someone else's choosing. When money can buy beliefs, especially beliefs that people themselves do not know were purchased, what does that mean for democracy? Will elections outcomes go to the highest bidder?

## 4 Recommendations: What could be done?

We have perhaps painted a bleak picture of the potential outcomes for BCI and Neurocapitalism. However, there are common themes in the scenarios that we have presented, and we believe there are things that can be done to mitigate their effects.

### 4.1 Education

One partial solution is improved education about the difference of the mental models that exist and about the rules that govern the digital space. Without such, the public may find difficulty in understanding what misconceptions they have with respect to how BCI technologies may affect them. An educated and skeptical public is the last, and also best defense against the dangers outlined above.

Scientists and Engineers within the field of BCI design can create works that communicate the capabilities and risks inherent in the use of BCIs. Science and Engineering communicators can make the understanding behind the technology of interest more engaging and help thwart such abuses. They can then hold forums with a more informed public to voice and channel concerns to scientists, artists, and engineers in research and development.

There are misconceptions about what constitutes privacy in the digital space, and what domains are public versus private. There is ignorance about what data can be collected, how it can be collected, how easily it can be copied, and how it will never decay, and how valuable it is. Instead of educating users about the value of their data, companies instead prefer to retain their information advantage over the user and coerce them into agreeing to unfavorable and predatory usage terms based on this ignorance.

Perhaps the greatest misconception is what constitutes you in the digital space. In the physical space we intuitively know what and where we are. We understand the value of our bodily safety, and our desire for privacy concerning information about our bodies is indicated by regulations medical records. Imagine a company whose business model was to collect hairs shed by people in public spaces with the intent of sequencing their genome, claiming that the hair was in a public space and therefore public domain, and that they now owned the rights to the sequenced genome because they spent the money to sequence it. If the idea of this makes you uneasy you are not alone, and genetic data is considered sensitive and understood to be integral to who we are. Insurance companies could use this as a means to systematically increase an individual's premiums. However, the digital equivalent of this scenario is currently playing out in the digital space. In the digital space, our body can be considered the sum total of information about us, especially our neural data. It should be granted similar protections in the digital space that our physical body has been in the physical space. A future without the requisite protections could take the shape of where totalitarian governments hunt those with aberrant thoughts or companies hunt those indefensibly susceptible to their products.

### 4.2 Ethical Design

The second line of defense is the engineers, scientists, and developers creating the hardware and software required to extract neural data from BCIs. Considering the current level of machine learning accuracy and the complexity of engineering problems facing BCI, there is a large burden of ethical responsibility that rests on those making these technologies, not only to consider the possible uses of the technology, but also to make certain their competency.

Machine learning models are rarely designed with the goal of being biased, or BCIs with the goal of incorrectly reading measurements, but both have occurred. For all the dangers that accompany a BCI-based personal behavior model, they are overshadowed by the dangers that come with a malfunctioning BCI-based personal behavior model. It is imperative that any such model, especially one that would be involved in deciding the fates of people, be properly designed, tested, and with the ability to be independently audited.

### 4.3 Regulation

Finally, there must be careful regulation and oversight of the companies dealing in personal data. The goal should not be to limit innovation, but limit how innovations can be used, in order to protect the populace. Companies cannot be fully trusted to act in the best interest of the populace, and often their motivations are at odds. Regulations are the voice of people and should represent their interests. A proper balance is both beneficial for both industry and the public good. There are many regulations which could mitigate or completely sidestep some of the issues mentioned above; they are the most direct form of intervention. Here we focus on a few areas where regulation might have the greatest effect. We accept that the use of technologies that allow increasing access to the mind cannot be stopped, but this path needs not be ruinous.

Companies are taking advantage of users with terms and conditions of unreasonable length and complexity. Owing to complex and drawn-out language, 91% of uses would blindly accept terms of agreement, whereas this percentage increased for users aged 18-34 in terms of willingness to even read the terms [2-4]. Another study found that terms coercing individuals to sign away valuables as high as their first born was relatively easy, which points at possible rights that many consumers have unknowingly been bereft of [2-4,16]. Regulations are required to outline and clarify the terms of agreement and other policy documents, making their wording less abstract and more accessible. This can ease consumer engagement, while also making them more knowledgeable about risks that are undertaken. A lack of openness would raise questions of if justice would be achieved with data acquired [16].

Another set of regulations that dovetail with the first two sections, are regulations that guide the research and development of BCIs and accompanying machine learning products. As BCIs fall within life science, they fall under Dual Use policy scopes and should be treated as such. As long as public funds are used, their development should be scrutinized and made transparent wherever possible to allow forums on use. This emphasis is further multiplied with concern to AI/ML algorithms being coupled with behavioral targeting.

Finally, regulations are needed to limit what can be done with user data and protect their right to privacy or ownership of that data. Currently, the data privacy and data management regulations of the European Union are the closest to what the authors would consider appropriate in the lengths that they go to protect users and the rights they afford. The argument can be made that they do not go far enough. We agree that data privacy should be protected with accountability, as well as deleted upon request. But in some scenarios, once information is given it is incorporated and can no longer be extricated. In these instances, in the commerce of neural data, there should be regulations for the fair compensation of the acquisition of such data. Users cannot demand a fair compensation if they are not operating with full information about the value of their data, which it is not in the best interest of the company to supply.

Finally, regulations are needed to limit what can be done with user data and protect their right to privacy or ownership of that data. Currently, the data privacy and data management regulations of the European Union are the closest to what the authors would consider appropriate in the lengths that they go to protect users and the rights they afford. The argument can be made that they do not go far enough. We agree that data privacy should be protected with accountability, as well as deleted upon request. But in some scenarios, once information is given it is incorporated and can no longer be extricated. In these instances, in the commerce of neural data, there should be regulations for the fair compensation of the acquisition of such data. Users cannot demand a fair compensation if they are not operating with full information about the value of their data, which it is not in the best interest of the company to supply.

These regulations are widely applicable to all aspects of the information economy and not limited to BCI. The earlier this segment of the economy is regulated, and users afforded greater protections, the surer will be the footing of the industry when BCI technology leaves the research lab. Beyond Neurocapitalism and BCI, everyone would benefit from an updated set of personal protections for the digital age. A digital Bill of Rights. From here, policy makers who are supported by a more informed and resilient public, can implement curated concerns balanced with industry into laws that overall improve and guide design, making for a more fertile market of development for society. The sooner meaningful regulation can be crafted, problems can be better identified sooner before it is a problem. Mass adoption of such technologies is still in the future.

## 5 Conclusions

The elements that we have discussed here are all local manifestations of systemic problems. Power seeks to consolidate. It will use any instrument available to do so. The digital space is a new frontier, not yet fully explored or understood. Maps are still being drawn and the borders of their territories negotiated. As is often the case, early arrivers have the information advantage. We have seen the Tech company giants parlay this information advantage into new hegemonies for the new space. The dangers that Neurocapitalism poses to people are the same as those of surveillance capitalism or any digital information commerce. Since information people produce about themselves is a commodity, companies have no incentive to educate them on the value of that data. Indeed, they have a strong economic incentive to keep people as ignorant as possible, of the value of their data, and the dangers that giving up the rights to it poses. Companies cannot be trusted to act in the best interest of people.

Governments must have stronger regulations in the digital space to protect people from exploitation, coercion, and control. The main barriers to this are the relative ignorance about the digital space, and the logistic complexity and novelty of regulating that space. The first can be addressed, while the second may always exist. Education is imperative to avoiding bad outcomes. People cannot ask for protections they do not know they need, and policymakers cannot regulate a problem that they do not understand or believe exists. These protections, while already imperative, become even more so with the addition of a BCI technology. If we are to have something as valuable as neural data bought and sold as a commodity, then we must take great care to severely limit the ways in which it can be used. If we do not take care to address improper development and implementation of BCI technology, we run the serious risk of handing a very effective new tool to those with disproportionate power to further expand and cement hegemonies.

BCI represents a new frontier, giving some access to the internal states of the mind. Such information is already highly sought after, so there is significant incentive to put resources towards the development of a technology that promises this kind of information. For this reason, we believe that it is only a matter of time before BCI technology becomes a consumer product.

We must prepare for its arrival and meet it on sure footing with proper protections already in place. If we are blindsided, or willfully ignorant of its potential abuses, the results could be disastrous to civil liberties. Surreptitiously influencing the hearts and minds of people is perhaps the most insidious form of control. Insidious because of its efficacy. We have discussed here how we believe this is already possible and stands to be made significantly more effective with the addition of BCI neural data. While beliefs, ideals, and opinions cannot be directly purchased, they can be influenced by the messages a person receives. Our way of life is determined by our shared beliefs as a society. In this way money is capable of directly steering our reality. If left unchecked, all this may add up to it being cheaper for the wealthy to buy the future than ever in our history. Are we comfortable with entities that we do not like having access to the keys to our means of communication? Will we craft new mental models for a new space that may emerge from our insights gained from this world of interaction and reading that we are embarking on?

In closing, we ask that humanity take a strong look at where it is going with Neurocapitalism and have the courage to ask tougher questions about its pace, and chart a new path, writing a new book.

## Acknowledgement

We would like to thank those who edited this with respect to their domains to make this a more digestible paper.